\documentclass[preprint2]{aastex}

\slugcomment{Draft v 2}

\shorttitle{A new low mass for the Hercules dSph}
\shortauthors{Ad\'{e}n et al.}

\begin{document}

\title{A new low mass for the Hercules dSph: the end of a common mass
  scale for the dwarfs?}

\author{D.\,Ad\'{e}n }
\affil{Lund Observatory, Box 43, SE-22100 Lund, Sweden}

\author{M.I.\,Wilkinson} \affil{Department of Physics and Astronomy,
  University of Leicester, University Road, Leicester LE1 7RH, UK}

\author{J.I.\,Read} \affil{Institute for Theoretical Physics,
  University of Zurich, Winterthurerstrasse 190 8047}
\affil{Department of Physics and Astronomy, University of Leicester,
  University Road, Leicester LE1 7RH, UK}

\author{S.\,Feltzing} \affil{Lund Observatory, Box 43, SE-22100 Lund,
  Sweden}

\author{A.\,Koch} \affil{Department of Physics and Astronomy,
  University of Leicester, University Road, Leicester LE1 7RH, UK}

\author{G.F.\,Gilmore} \affil{Institute of Astronomy, Madingley Road,
  Cambridge, CB3 0HA, UK}

\author{E.K.\,Grebel} \affil{Astronomisches Rechen-Institut, Zentrum
  f\"ur Astronomie der Universit\"at Heidelberg, M\"onchhofstr. 12-14,
  69120 Heidelberg, Germany}

\author{I.\,Lundstr\"om} \affil{Lund Observatory, Box 43, SE-22100
  Lund, Sweden}

\begin{abstract}
  We present a new mass estimate for the Hercules dwarf spheroidal
  galaxy (dSph), based on the revised velocity dispersion obtained by
  \citet{Aden2009}.

  The removal of a significant foreground contamination using newly
  acquired Str\"omgren photometry has resulted in a reduced velocity
  dispersion. Using this new velocity dispersion of $3.72 \pm 0.91
  {\rm \, km\, s^{-1}}$, we find a mass of $M_{300}=1.9^{+1.1}_{-0.8}
  \times\, 10^6 \,M_{\odot}$ within the central 300\,pc, which is also
  the half-light radius, and a mass of
  $M_{433}=3.7_{-1.6}^{+2.2}\times10^6$M$_\odot$ within the reach of
  our data to 433\,pc, significantly lower than previous estimates. We
  derive an overall mass-to-light ratio of $M_{433}/L=103^{+83}_{-48}
  \, [M_{\odot}/L_{\odot}]$. Our mass estimate calls into question
  recent claims of a common mass scale for dSph galaxies.

  Additionally, we find tentative evidence for a velocity gradient in
  our kinematic data of $16 \pm 3$\,km\,s$^{-1}$kpc$^{-1}$, and
  evidence of an asymmetric extension in the light distribution at
  $\sim 0.5$\,kpc. We explore the possibility that these features are
  due to tidal interactions with the Milky Way.  We show that there is
  a self-consistent model in which Hercules has an assumed tidal
  radius of $r_t = 485$\,pc, an orbital pericentre of $r_p = 18.5\pm
  5$\,kpc, and a mass within $r_t$ of
  $M_{\mathrm{tid},r_t}=5.2_{-2.7}^{+2.7} \times
  10^6$\,M$_\odot$. Proper motions are required to test this
  model. Although we cannot exclude models in which Hercules contains
  no dark matter, we argue that Hercules is more likely to be a dark
  matter dominated system which is currently experiencing some tidal
  disturbance of its outer parts.

\end{abstract}


\section{Introduction}

Dwarf spheroidal (dSph) galaxies are believed to play an important
role in the formation and evolution of much more luminous galaxies
\citep[e.g.][]{1994PASP..106.1225G}. dSphs are characterized by their
low surface brightness, low total luminosity and spheroidal shapes
which are consistent with their pressure-supported stellar
kinematics~\citep[e.g.][]{Grebel2003}.

Knowledge of dSph masses is essential for comparison with cosmological
simulations of galaxy formation. Good mass estimates help us to
establish whether the paucity in the number of observed systems (a few
tens) to the number of predicted satellite haloes (several thousands)
represents a fundamental failure of our cosmological model
\citep[e.g.][]{1999ApJ...524L..19M}, or whether it is simply telling
us that galaxy formation is inefficient on small scales
\citep[e.g.][]{2006MNRAS.371..885R}. Recent studies have suggested
that the dSph galaxies share a common mass within a certain radius
\citep{2007ApJ...667L..53W,2008Natur.454.1096S,2009arXiv0906.0341W}.
If confirmed, this would be an important clue to the processes which
regulate the formation of the lowest luminosity galaxies.

All mass estimates implicitly assume that contamination of the
kinematic sample by foreground stars or unbound tidal tails is
negligible, and that the system is in (or close to) virial
equilibrium.  Since the mass of an equilibrium stellar system is
proportional to its velocity dispersion squared, an over-estimate of
the velocity dispersion will result in an inflated mass estimate.  The
assumption of virial equilibrium may be called into question if the
system is tidally disrupting
~\citep{Oh1995,Kroupa1997,2007MNRAS.378..353K,2008ApJ...679..346M}.

The recently discovered Hercules dSph~\citep{Belokurov2007}, with an
ellipticity of $e\sim 0.5$ \citep{2007ApJ...668L..43C}, is an example
of a galaxy in which assumptions of equilibrium may be incorrect.  In
\citet{Aden2009} we showed that the mean velocity of the Hercules dSph
is embedded in the foreground dwarf star velocity distribution. In
order to weed out the foreground dwarf stars we used the Str\"omgren
$c_1$ index. This index is able to clearly identify red-giant branch
(RGB) stars redder than the horizontal branch, enabling a separation
of RGB stars in the dSph galaxy and foreground dwarf stars.  By
weeding out the foreground contaminants we found that the dispersion
for Hercules is reduced from $7.33 \pm 1.08 {\rm \, km\, s^{-1}}$ to
$3.72 \pm 0.91 {\rm \, km\, s^{-1}}$.  In this letter we explore the
consequences of this finding.

In Section~2, we derive a new mass for Hercules and show that it is
not consistent with a common mass scale for the dSphs. In Section~3,
we investigate the possibility of a velocity gradient in the kinematic
data of Hercules. In Section~4, we discuss the relative importance of
tides and dark matter in Hercules.  Section~5 summarises our
conclusions.

\section{Mass estimate from the spherical Jeans
  equations} \label{mass}

We use the velocity dispersion, $\sigma_v=3.72 \pm 0.91 {\rm \, km\,
  s^{-1}}$ \citep{Aden2009}, to constrain the mass of the Hercules
dSph. We assume that the system is in dynamical equilibrium, is
spherically symmetric, has an isotropic velocity distribution, and a
flat velocity dispersion profile\footnote{Note that
  \citet{2009arXiv0908.2995W} and \citet{2009arXiv0906.0341W} find
  that their mass estimates at the half light radius are insensitive
  to a wide range of mass models and velocity anisotropy
  parameterisations, so our results should not be sensitive to these
  assumptions.}.  With these assumptions, the Jeans equation for the
mass distribution \citep[Eq. (4.215) in][]{2008gady.book.....B}
becomes:
\begin{equation} \label{eq1} \sigma_v^2 \, \frac{d\,
    \nu(r)}{dr}=-\frac{\nu(r)\, G\, M(r)}{r^2}
\end{equation}
where $r$ is the three-dimensional distance from the centre of the
galaxy, $\nu(r)$ is the de-projected stellar density profile and
$M(r)$ is the enclosed mass.  We use the exponential profile from
\citet{2008ApJ...684.1075M} to describe the stellar density
profile. The de-projected exponential profile is given by
\citet{2007MNRAS.378..353K} as (setting $m=1$ in their Eq. (4)):
\begin{equation} \label{eq2} \nu(r)=\nu_0\left( \frac{r}{\alpha}
  \right)^{-0.445}\, e^{-r/\alpha}
\end{equation}
where $\alpha$ is the exponential scale radius, related to the
half-light radius as $r_h=1.68\, \alpha$, and $\nu_0$ is the central
stellar density. Solving Eq. (\ref{eq1}) for $M(r)$ using
Eq. (\ref{eq2}) yields
\begin{equation} \label{eq3} M(r)=\frac{r(r+0.445 \,
    \alpha)\sigma_v^2}{\alpha \, G}
\end{equation}
Using a half-light radius of $330^{+75}_{-52}$\,pc
\citep{2008ApJ...684.1075M} we solve Eq. (\ref{eq3}) for $r=433$ pc,
which corresponds to the outermost member in our kinematic sample
(Table \ref{tab}).

\begin{figure}
\epsscale{.90}
\plotone{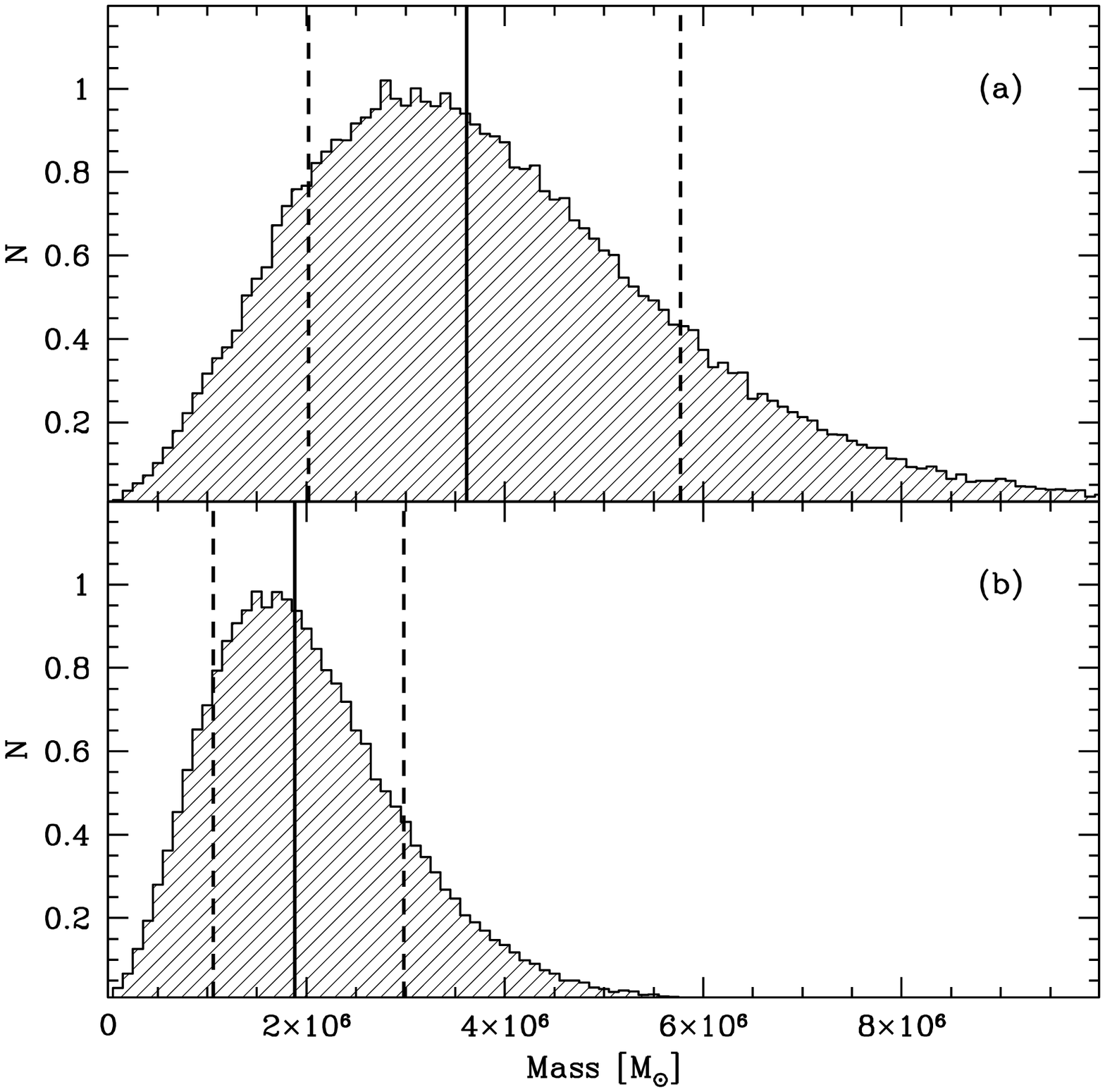}
\caption{Distribution of Monte Carlo mass estimates for
  Hercules. Solid line indicates the median from the Monte Carlo
  sampling. Dashed lines indicate the $1\sigma$ limits from the Monte
  Carlo sampling. ({\bf a}) Mass within the radius defined by the
  outermost RGB star in the kinematic sample, $r=433$ pc. ({\bf b})
  Mass within the inner 300 pc. \label{fig1}}
\end{figure}

We estimate the error in the mass using $10^5$ Monte Carlo
realisations of the half-light radius, distance and velocity
dispersion of Hercules drawn from within the individual error bars on
each parameter. In this way, we obtain $M_{433}=3.7^{+2.2}_{-1.6}
\times \, 10^6 \, M_{\odot}$ (Fig. \ref{fig1}).  The quoted errors are
1-$\sigma$ limits from the Monte Carlo sampling.  Assuming a total
luminosity of $L=(3.6 \pm 1.1) \times 10^4 \, L_{\odot}$
\citep{2008ApJ...684.1075M}, we find a median mass-to-light ratio of
$M_{433}/L=103^{+83}_{-48} \, [M_{\odot}/L_{\odot}]$.

\citet{Aden2009} emphasise the importance of weeding out foreground,
contaminating dwarf stars. This is particularly vital for the Hercules
dSph as its systemic velocity coincides with the bulk motion of dwarf
stars in the Milky Way disk in the direction of Hercules. Using the
contaminated\footnote{Note that this ``contaminated" dispersion has
  had its 3$\sigma$ velocity outliers removed. Also, the initial
  candidate selection for the velocities were chosen using
  colour-magnitude diagram cuts. Even so, a significant fraction of
  foreground stars remain in this sample and are detected only through
  the use of the Str\"omgren photometry.} velocity dispersion (Table
\ref{tab}) we would obtain a mass of $1.4^{+0.5}_{-0.4} \times 10^7 \,
M_{\odot}$, almost a factor of four larger than the uncontaminated
estimate.

\subsection{Hercules and the ``Common Mass Scale''}

\citet{2008Natur.454.1096S} speculate that the mass within a fixed
radius, 300\,pc, is approximately the same ($\sim 10^7 M_{\odot}$) in
all the observed dSphs. Our revised velocity dispersion implies a mass
within the inner 300 pc (M$_{300}$) of only $1.9^{+1.1}_{-0.8}\times
\, 10^6 \, M_{\odot}$. This indicates that Hercules falls considerably
below this ``common mass scale'' for dSphs.  Since 300 pc is also
approximately the half-light radius, this implies that Hercules also
lies significantly off the \citet{2009arXiv0906.0341W} enclosed
half-light mass scaling relation.

To confirm that our result is not sensitive to our choice of surface
brightness profile, we have repeated our calculation using a Plummer
profile with scale radius $r_{pl}=321 {\rm pc}$~\citep[as used
by][]{2008Natur.454.1096S}.  In this case we find M$_{300} = 2.3
\times \, 10^6 \, M_{\odot}$ in agreement with the mass calculated
above. If instead we use our contaminated velocity dispersion and the
exponential surface density profile, we obtain M$_{300} =
7.4^{+2.7}_{-2.1} \times \, 10^6 \, M_{\odot}$ which agrees with the
value presented in \citet{2008Natur.454.1096S}.

However, we note that \citet{2008Natur.454.1096S} used a dispersion of
$5.1\pm 0.9{\rm\,km\,s^{-1}}$ \citep[taken
from][]{2007ApJ...670..313S} to obtain their mass estimate. This is
smaller than our contaminated value of $7.33 \pm 1.08{\rm\,km
  \,s^{-1}}$, yet they obtain a similar median mass to us. If we use a
dispersion of $5.1 \pm 0.9{\rm\,km \,s^{-1}}$ as they do, we obtain a
mass $M_{300}=3.6^{+1.5}_{-1.2} \times 10^6$\,M$_\odot$. This is just
consistent with their determination within our mutual error bars. We
note that the median $M_{300}$ will depend on the details of the mass
modelling procedure and so can be expected to differ between their
study and ours (Matt Walker priv. comm.).

Additionally, we have repeated our calculation of the mass within
300\,pc using a more recent estimate of the half-light radius,
$r_h=230$\,pc, by \citet{2009arXiv0906.4017S}.  This half-light radius
is smaller and gives a mass that is $\sim 1.5$ times more massive than
the mass calculated using the half-light radius from
\citet{2008ApJ...684.1075M}.

\section{A velocity gradient in Hercules}

The presence of a velocity gradient in a dSph could either be
indicative of an intrinsic rotation, or a sign of tidal interaction
with the Milky Way. In this section, we test for possible velocity
gradients in Hercules.

\begin{figure}
\epsscale{.90}
\plotone{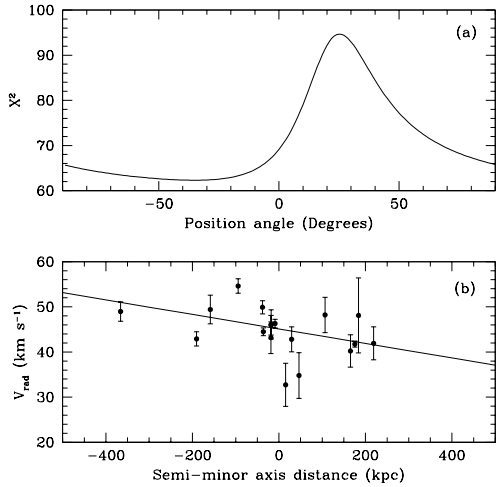}
\caption{{\bf (a)} $ \chi^2 $ vs. the position angle. {\bf (b)}
  $V_{rad}$ vs. semi-minor axis distance. The solid line indicates the
  least-square fit. $\bullet$ indicates members based on Str\"omgren
  photometry \citep[see][]{Aden2009}.}
    \label{fig2}
\end{figure}

Assuming that the rotation around the semi-minor axis (in the ellipse
that desribes the orientation of the dSph) is more likely then around
the semi-major axis we can define the "semi-minor axis distance"
$d_{{\rm mi}}(\theta)$ as the distance, perpendicular to the
semi-minor axis, between the axis and the star.  For each of the 18
RGB stars with radial velocities we calculate this distance for
different position angles $\theta$. We then derive the gradient ,
$k_{\rm rot}$, and zeropoint, $m_v$, by solving the following equation
with a least-squares fit for each position angle,
\begin{equation}
  V_{{\rm rad}}=k_{\rm rot}\times d_{{\rm mi}}(\theta)+m_v
\end{equation}
where $V_{{\rm rad}}$ is the radial velocity measurement for each
stars.  The least-squares fit to this function yields a $\chi^2$ value
for each $\theta$ (Fig. \ref{fig2}a).

The distribution enclosed by $\chi^2_{\rm min}+1$ corresponds to $1 \,
\sigma$ for a normal distribution \citep{1992nrfa.book.....P}. We use
this to obtain the error in the position angle which minimizes
$\chi^2$.

We find a position angle of $-35^{+18}_{-23}$ degrees with a velocity
gradient of $16 \pm 3$\,km\,s$^{-1}$kpc$^{-1}$, and a zeropoint of
$45.11\pm 0.38{\rm \, km\, s^{-1}}$. We obtain a reduced $\chi^2$ of
$3.89$ for our 18 stars with 2 degrees of freedom.

Following \citet{2008ApJ...688L..75W}, we estimate the significance of
the velocity gradient using $10^5$ Monte Carlo realisations. In each
realisation, we sample the velocity and spatial distributions
independently. Thus we scramble the correlation between velocity and
position, while maintaining the original velocity distribution and
spatial positions. This is valid as long as the phase space
distribution function of the stars is separable (which implies that
the velocity dispersion is independent of radius). We determine the
significance of the velocity gradient by computing the fraction of
realisations that fail to produce a $\chi^2$ as low as the one
calculated from the real data. We find a significance of the velocity
gradient of 78 per cent ($1.23 \sigma$).

\section{Discussion}
\subsection{Galactic tides in Hercules?}

\begin{figure}
\epsscale{.90}
\plotone{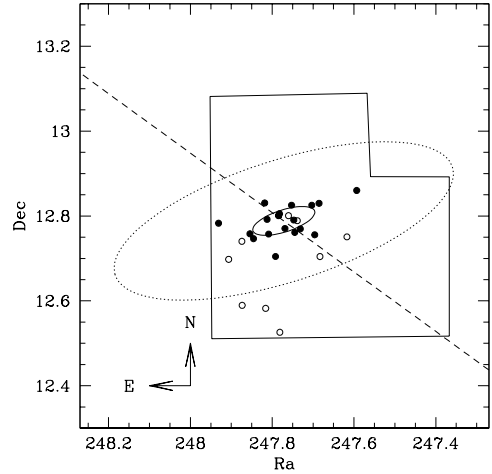}
\caption{Spatial distribution on the sky of the stars identified as
  Hercules members from both spectroscopy and photometry. $\circ$
  indicate members based on Str\"omgren photometry. $\bullet$ indicate
  members based on Str\"omgren photometry for which there are radial
  velocity measurements.  The dashed line indicates the semi-minor
  axis for the position angle of our detected rotation axis.  The
  solid ellipse represents the core radius and the dotted ellipse the
  inner border of the field region of the Hercules dSph
  \protect\citep{2007ApJ...668L..43C}. Solid lines outline the
  footprint of the WFC used to obtain the Str\"omgren
  photometry. \label{fig3}}
\end{figure}

In the previous section we found tentative evidence for a velocity
gradient in the Hercules dSph. The peak-to-peak difference of $\sim 10
\, {\rm km \, s^{-1}}$ within a radius of less than 1\,kpc could be
interpreted as the effect of Galactic tides
\citep{2006MNRAS.367..387R}.  Interestingly, \citet{Martin2009}
recently estimated the orbit of Hercules based on the assumption that
its elongation is tidally induced, and predicted a velocity gradient
of at least $7\, {\rm km \, s^{-1}} {\rm kpc}^{-1}$ which is
consistent with our observed gradient.  Additionally, in
\citet{Aden2009} we found that the spatial distribution of the
Hercules stars is asymmetric at $\sim$0.5 kpc.  There are three
significant outlier stars to the South, but no corresponding members
at this distance in either the Northern or Western fields
(Fig. \ref{fig3}).  These three stars are unambiguously identified as
RGB stars from Str\"omgren photometry.  We now consider the
possibility that the velocity gradient and the positional outliers are
evidence that Hercules is being tidally disrupted, and use this
information to obtain a second mass estimate for the system.

The tidal radius of a dSph depends on the potential of the host
galaxy, the potential of the dSph, the orbit of the dSph within the
host galaxy and the orbit of the stars within the dSph
\citep[e.g.][]{2006MNRAS.366..429R}.  We parameterise the Milky Way
potential using the default model in \citet{2005ApJ...619..800J},
analysed in the Galactic plane; and the Hercules potential using a
generalised Hernquist profile \citep{1990ApJ...356..359H}:

\begin{equation}
  \rho(r) = \frac{M(3-\gamma)}{4\pi r_s^3}\left(\frac{r}{r_s}\right
  )^{-\gamma}\left(1+\frac{r}{r_s}\right)^{\gamma-4}
\label{eqn:hernquist}
\end{equation}
where $M$ is the total mass, $r_s$ the scalelength, and $\gamma$ the
central logarithmic cusp slope. We consider the ranges
$0.3<r_s<3$\,kpc and $0<\gamma<1$. Our results are not sensitive to
these choices.

The orbit of the Hercules dSph is unknown but it currently lies at a
distance of $132$\,kpc from the galactic centre, and has a
heliocentric velocity of $45.2 \pm 1.09 {\rm \, km\, s^{-1}}$, which
implies a galactocentric radial velocity of $145 \pm 1.09 {\rm \, km\,
  s^{-1}}$ \citep[see Eq. (5) in][]{1999AJ....118..337C}. Using the
above potential model for the Milky Way and setting the tangential
velocity component for Hercules to zero, this gives us a minimum
apocentre for Hercules of $r_{a,\mathrm{min}} = 188.5$\,kpc.  Thus, we
consider apocentre and pericentre ranges of $188.5 < r_a < 600$\,kpc
and $5< r_p<132$\,kpc, respectively.

\begin{figure}
\epsscale{.90}
\plotone{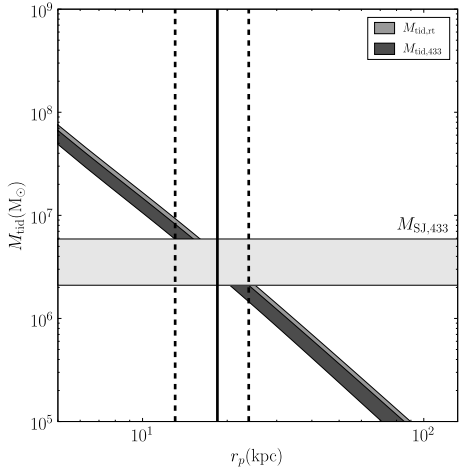}
\caption{Tidally determined mass for Hercules as a function of orbital
  pericentre $r_p$. We assume a retrograde tidal radius of $r_t =
  1$\,kpc for this calculation. The horizontal grey band marks our
  mass estimate from the spherical Jeans equation $M_\mathrm{SJ,433}$
  (See Sect. 2). The dark grey band marks our tidally determined mass
  within 433\,pc $M_\mathrm{tid,433}$ for Hercules. The medium grey
  band shows $M_{\mathrm{tid},r_t}$, our tidally determined mass
  within $r_t$. The vertical solid and dashed lines mark the
  pericentres at which $M_\mathrm{tid,433} = M_\mathrm{SJ,433}$ (see
  Eq. \ref{eqn:masseq} for a formula giving the scaling of
  $M_{\mathrm{tid},r_t}$ with $r_t$).}
\label{fig4}
\end{figure}

\citet{2006MNRAS.366..429R} find that photometric features are
typically seen beyond the retrograde tidal radius.  To proceed, we
make the assumption that the outliers in Fig. \ref{fig3} indicate the
location of the retrograde tidal radius of Hercules, i.e. $r_t \sim
485$\,pc. If we assume further that the tidal radius of Hercules is
set at the pericentre of its orbit, we can solve Eq. (7) of
\citet{2006MNRAS.366..429R} to calculate the mass, $M_{\mathrm{tid}}$,
of Hercules as a function of its orbital pericentre $r_p$, for our
assumed ranges of $r_s$, $\gamma$ and $r_a$. We calculate both its
mass within $r_t$ ($M_{\mathrm{tid},r_t}$) and its mass within 433\,pc
($M_\mathrm{tid,433}$) which can then be compared with our mass
estimate based on the spherical Jeans equation
($M_{\mathrm{SJ},433}$). The results of this calculation are given in
Fig. \ref{fig4}. The horizontal light grey band marks
$M_{\mathrm{SJ},433}$, the dark grey band marks $M_\mathrm{tid,433}$
and the medium grey band marks $M_{\mathrm{tid},r_t}$. The vertical
solid and dashed lines mark the pericentres at which
$M_\mathrm{tid,433}=M_\mathrm{SJ,433}$. The width of the tidal mass
bands is due to the unknowns: $r_s$, $\gamma$, and $r_a$. However, to
lowest order $r_t$ depends only on the mean density enclosed within
it, and so these bands are narrow. For this reason, we obtain an
estimate of both the orbital pericentre of Hercules ($r_p = 18.5 \pm
5$\,kpc) and its mass within the tidal radius
($M_{\mathrm{tid},r_t}=5.2_{-2.7}^{+2.7} \times 10^6$M$_\odot$). The
primary source of error on both of these quantities is our assumed
tidal radius $r_t$. Empirically we derive scalings of:

\begin{equation}
r_p = 32 \left(\frac{r_t}{1\,\mathrm{kpc}}\right)^\beta\,\mathrm{kpc}
\label{eqn:masseq}
\end{equation}
with $\beta = 0.76$ over the range $r_t = [0.4,2]$\,kpc, and: 

\begin{equation}
  M_{\mathrm{tid},r_t}=24 \times 10^6 \left(\frac{r_t}{1\,\mathrm{kpc}}
  \right)^\alpha\,\mathrm{M}_\odot 
\label{eqn:rpeq}
\end{equation}
with $\alpha = 2.1$ over the same range.

If we assume $r_{\rm t}=485$ pc, we obtain a similar estimate of the
pericentric distance as that obtained by \citet{Martin2009}.

\subsection{Is Hercules a dark matter free system?}  \label{disc5}

We have shown that our Hercules data are consistent with the presence
of dark matter. We now consider whether models without dark matter
could also reproduce the data.

The most extreme scenario is that Hercules is disintegrating and its
velocity dispersion arises solely due to the motion of its unbound
member stars. If so, it will rapidly become unobservable (i.e. reach a
lower surface brightness than the detection limit of the
SDSS). Hercules has a surface brightness of $27.2\pm0.6$ mag
arcsec$^{-2}$ \citep{2008ApJ...684.1075M}. If the unbound stars are
moving away from Hercules at a velocity equal to the velocity
dispersion ($3.72 \pm 0.91 {\rm \, km\, s^{-1}}$) it would require
only $\sim 200 \times 10^6$ years for the Hercules dSph to fall below
the detection limit of SDSS~\citep[$\sim 30 {\rm \, mag \,
  arcsec^{-2}}$][]{2008ApJ...686..279K}. Given this short timescale,
it is very unlikely that we would observe Hercules at this phase of
its evolution.

\citet{2007MNRAS.375.1171F} simulated the disruption of the UMa II
dSph.  Using a model that does not distinguish between luminous and
dark matter they simulate the surface brightness, velocity dispersion
and the mean radial velocity for UMa II after 9, 10 and 11\,Gyr (their
Fig. 9).

In the absence of dark matter, the equilibrium velocity dispersion
(i.e. $\sigma^2=G\, M/r$) for the Hercules dSph would be $\sim 1 {\rm
  \, km\, s^{-1}}$, assuming a stellar mass of $5 \times 10^4 \, {\rm
  M_{\odot}}$ and a radius equal to the observed half-light radius
\citep[$300$\,pc;][]{2008ApJ...684.1075M}. This is much lower than the
measured velocity dispersion of $3.72 \pm 0.91 {\rm \, km\,
  s^{-1}}$. We conclude that the Hercules dSph can only be a dark
matter free system if its velocity dispersion has been inflated
significantly - i.e. if it is in the advanced stages of tidal
disruption.

However, in the simulations of \citet{2007MNRAS.375.1171F} the
evolutionary phase in which a tidal remnant exhibits both an inflated
velocity dispersion and a velocity gradient, and remains centrally
concentrated is very short. Within $\sim 1$\,Gyr the system goes from
bound, with almost no signs of tidal disturbance, to complete
disruption.

(We note that \cite{Kroupa1997} performed simulations of dSph galaxies
without dark matter, and found that it is possible to obtain
long-lived remnants whose properties are remarkably similar to those
of Hercules in terms of velocity dispersion and velocity gradient (see
his Fig. 9). However, the remnants simulated by \cite{Kroupa1997} are
significantly more luminous than the Hercules dSph. Based on the
simulations of \citet{2007MNRAS.375.1171F}, it seems unlikely that
lower luminosity, purely stellar remnants with the correct properties
could survive for a significant time.)

It seems difficult to understand Hercules without dark
matter. However, it could be dark matter dominated {\it and}
experiencing significant tidal disturbance. In this case, our simple
Jeans analysis may have over-estimated its mass and Hercules might lie
even further away from a possible common mass scale for the dSph
galaxies.

\section{Conclusions}
We have calculated the mass of the Hercules dSph using the new
velocity dispersion for the system obtained by \citet{Aden2009}. We
find that the mass within the volume enclosed by our observed stars is
$3.7^{+2.2}_{-1.6}\times \, 10^6 \, M_{\odot}$, leading to a
mass-to-light ratio of $103^{+83}_{-48} \, [M_{\odot}/L_{\odot}]$.
Interestingly, the mass within 300\,pc is significantly lower than the
``common mass scale'' found by \citet{2008Natur.454.1096S}, suggesting
that Hercules does not share the halo properties seen in other dSphs.

We found tentative evidence for a velocity gradient of $16 \pm
3$\,km\,s$^{-1}$kpc$^{-1}$, and evidence of an asymmetric extension in
the light distribution at $\sim0.5$\,kpc.  We explored the hypothesis
that these features are due to tidal interactions with the Milky Way.
Assuming a tidal radius of $485$\,pc, we show that a self-consistent
model requires Hercules to be on an orbit with pericentre $r_p = 18.5
\pm 5$\,kpc, and with a mass within $r_t$ of
$M_{\mathrm{tid},r_t}=5.2_{-2.7}^{+2.7}\times10^6$M$_\odot$.

\acknowledgments
 
We are very grateful to Matt Walker for useful discussions regarding
the mass estimates in earlier works. We thank Pavel Kroupa for
stimulating discussions.  D.A thanks Lennart Lindegren at Lund
Observatory for help with statistics.  S.F. is a Royal Swedish Academy
of Sciences Research Fellow supported by a grant from the Knut and
Alice Wallenberg Foundation. M.I.W is supported by a Royal Society
University Research Fellowship.

\clearpage

\clearpage

\begin{deluxetable}{l l c c c c c c}
  \tabletypesize{\scriptsize} \rotate \tablecaption{Summary of
    determination of systemic velocities, velocity dispersions, masses
    and metallicities for the Hercules dSph.}  \tablewidth{0pt}
  \tablehead{ \\
    \colhead{} & \colhead{Number of stars} & \colhead{$v_{sys}$} &
    \colhead{$\sigma$} & \colhead{${\rm M}_{ \rm Furthest \, \,
        star}$} & \colhead{M$_{300 pc}$} &
    \colhead{$\langle {\rm [Fe/H]} \rangle $}\tablenotemark{a}  \\

    \colhead{} & \colhead{} & \colhead{ [${\rm \, km\, s^{-1}}$]} &
    \colhead{ [${\rm \, km\, s^{-1}}$]} & \colhead{[$M_{\odot}$]} &
    \colhead{[$M_{\odot}$]} &
    \colhead{dex}  \\
  } \startdata
  \citet{Aden2009} & 28 [RGB stars, $c_1$ sel.] & - & - & - & - & $-2.35\pm 0.31$  \\
  This study &  32 [Only $V_{rad}$ sel.] & $40.87\pm 1.42$ & $7.33 \pm 1.08$ & $1.4^{+0.5}_{-0.4} \times \, 10^7$ & $7.4^{+2.7}_{-2.1} \times \, 10^6$ & - \\
  This study & 18 [RGB stars with $V_{rad}$] & $45.20 \pm 1.09$ & $3.72 \pm 0.91$ & $3.7^{+2.2}_{-1.6} \times \, 10^6$ & $1.9^{+1.1}_{-0.8} \times \, 10^6$ & -  \\
\enddata
\tablenotetext{a}{See \citet{Aden2009}}

\tablecomments{Columns: (1,2) number of stars in each study and a
  short description of how the stars were selected; (3,4) systemic
  velocities and velocity dispersions; (5) mass within the radius
  defined by the outermost RGB star in our study; (6) mass within the
  inner 300 pc; (7) mean metallicity [see \citet{Aden2009} for a full
  discussion of metallicities].}
\label{tab}
\end{deluxetable}

\end{document}